\documentclass[prl,superscriptaddress,showpacs,twocolumn]{revtex4}

\usepackage{amsmath}
\usepackage{amssymb}

\begin{document}

\title{ 
Gauge theory of topological phases of matter
}

\begin{abstract}
We study the response of quantum many-body systems to coupling some of their degrees of freedom to external gauge fields. This serves to understand the current Green functions and transport properties of interacting many-body systems. Our analysis leads to a ``gauge theory of states of matter" complementary to the well known Landau theory of order parameters.
We illustrate the power of our approach by deriving and interpreting the gauge-invariant effective actions of (topological) superconductors, 2D electron gases exhibiting the quantized Hall- and spin-Hall effect, 3D topological insulators, as well as axion electrodynamics. We also use the theory to elucidate the structure of surface modes in these systems.
\end{abstract}

\author{J\"urg Fr\"ohlich
}
\affiliation{Theoretische Physik, ETH Zurich, 8093 Z{\"u}rich, Switzerland}
\altaffiliation[Present address: ]{School of Mathematics, The Institute for Advanced Study, Princeton, NJ 08540, USA}
\author{Philipp Werner}
\affiliation{Department of Physics, University of Fribourg, 1700 Fribourg, Switzerland}


\maketitle

During the past several years, fascinating
novel states of condensed matter protected by topological properties 
(but not describable by local order parameters) have been predicted and found in experiments; 
see Refs.~\cite{Kane2005, Bernevig2006, Koenig2007, Koenig2008, Fu2007, Ludwig, 
Hsieh2008, Kane_rmp, Zhang_rmp, SS} and references given therein.
The purpose of this Letter is to sketch an approach towards classifying states of condensed matter, appropriately called \textit{``gauge theory of states of matter"}, that is complementary to the well known \textit{Landau theory of order parameters}. It can be used to study states of systems of condensed matter exhibiting ``topological order''. Its main field of application lies in the study of low-temperature properties of systems with a bulk energy gap, i.e., \textit{(topological) insulators} and \textit{incompressible quantum liquids}. For this class of systems, our theory makes precise predictions of response laws and transport equations and of the structure of surface states. The main ideas underlying our approach can also be used in the study of superconductors \cite{S&K}, cold atom gases \cite{Spielmanetal}, etc. 

The principal contours of a ``gauge theory of states of matter'' have been developed in the 90's \cite{FS, FST, FP, Alekseev1998, Werner2000}. Here we show how a suitable extension of this theory can be used to interpret various recent theoretical predictions and experimental findings in a novel way, independently of the interaction strength in the system; (for earlier work, see also Refs.~\cite{Anandan, Haldane, Prange-Girvin, Wen, FK, FGoetsch, FKST}). 
While our theory predicts the general form 
of the equations governing the transport in systems with a bulk energy gap, 
it does usually \textit{not} yield quantitative predictions for the values of various transport coefficients, such as conductivities, dielectric tensors or magnetic permeabilities. For quantized transport coefficients of incompressible electron liquids or of ``non-trivial band insulators'', such predictions can sometimes be inferred from ``$k$-space topology" (Chern numbers of complex vector bundles constructed from one-particle wave functions over $k$-space, indices, etc.) \cite{TKNN, BES, ASS, Kane2005, Bernevig2006, Fu2007, Ludwig, Hsieh2008, Kane_rmp, Zhang_rmp, Grafetal}, or from topological field theories arising as effective low-energy theories of such systems \cite{FST, Alekseev1998, FKST, Toappear}. The methods developed in these references are complementary to those described in this Letter and tend to have a more limited range of applicability than the approach presented here. 

The purpose of this Letter is to present a short, non-technical account of the main features and new applications of our theory. 
A more technical discussion will appear in \cite{Toappear}; see also \cite{FST}.
We will explicitly consider \textit{electron gases}, but the theory also applies to other systems, such as cold atom gases \cite{FST, Volovik}, the primordial plasma in the universe \cite{BoyFrRuch}, etc. Our \textit{main new results} concern (i) the structure of edge spin currents in certain quantum Hall fluids and in a variant of graphene with a bulk mobility gap, (ii) the structure of surface modes and the general form of the effective action of 3D topological insulators and superconductors, and (iii) a class of magnetic topological insulators with domain walls carrying extended states, which are described by axion electrodynamics. Besides reporting on these new insights, we recall some important findings presented in earlier papers, such as the general theory of the spin-Hall effect (in band insulators with unbroken time-reversal invariance) \cite{FS,FST} that has attracted much attention, during the past several years.

The key idea underlying our theory is to analyze states of electron gases by studying their response to turning on various external gauge fields. The Pauli equation of a non-relativistic, spinning electron moving in an external potential reads
$i\hbar \partial_t \Psi_t = H_t \Psi_t,$ where the Hamiltonian is given by $H_t=\frac{1}{2m}[\frac{\hbar}{i}\vec \nabla]^2+V(x,t)$, $V$ is the potential and $\Psi_t(x)$ a two-component spinor. This equation is invariant under global phase- and $SU(2)$-transformations of $\Psi_t$.  In accordance with Noether's theorem, there are two conserved current densities, the electric- and the spin current density. We promote these global symmetries to \textit{local} gauge symmetries by introducing $U(1)$- and $SU(2)$-gauge fields (vector potentials), $a$ and $w$. In many applications, these gauge fields describe internal or external electromagnetic fields, spin-orbit- and Zeeman interactions, exchange interactions (in magnetic materials), effects due to the curvature of the sample (as well as dislocations and disclinations), and the influence on electronic properties of the motion of the ionic background harboring the electron gas. They may be ``intrinsic'' and need not be small. In other applications, they are ``fictitious'', in the sense that they do not correspond to actual gauge fields acting on the system, but are introduced for the sole purpose of deriving response laws.
In order to take into account effects of the motion of the sample harboring the electron gas, the Pauli equation can be written in moving coordinates corresponding to the flow generated by a (divergence-free) velocity field $\vec v$, which yields contributions to the gauge fields 
$a$ and $w$, as described below. 

The resulting Pauli Hamiltonian is given by
\begin{equation}
H_t=\frac{1}{2m}\left[ \frac{\hbar}{i}\vec\nabla+\vec a+\vec w^K\sigma_K\right]^2+a_0+w^K_0\sigma_K.
\end{equation}
Here, $\sigma_K$, $K=1,2,3$, are the usual Pauli matrices; $\vec a(x,t)= \vec A^{(0)}(x,t) + \frac{e}{c}\vec A^\text{em}(x,t)+m\vec v(x,t)$, where $\vec A^{(0)}$ may describe a static, homogeneous external electromagnetic field and/or geometrical properties of the sample, $\vec A^\text{em}$ is the vector potential of fluctuations in the electromagnetic field, and $\vec v$ is the velocity field describing the motion of the ionic background \cite{footnote1}. 
The constant $e$ is the elementary electric  charge, $c$ the speed of light, and $m$ the electron mass.
The potential $a_0$ is given by $a_0(x,t)=\frac{e}{c}\varphi(x,t)-\frac{m}{2}\vec {v}^{2}(x,t) +\frac{1}{\rho}p(x,t)$, where $\varphi$ is the electrostatic potential, $p$ the pressure and $\rho$ the density of the ionic background. The $SU(2)$-gauge field, $w$, is  
\begin{align}
\vec w^K \sigma_K &= \vec\Pi \wedge \frac{\hbar\vec \sigma}{2}+\vec W^K \sigma_K+\vec\Omega^K\sigma_K,\label{W1}\\
w_0^K \sigma_K &= \frac{\mu}{\hbar}(\vec B + \frac{mc}{e}\vec \nabla \wedge \vec v ) \cdot \frac{\hbar\vec \sigma}{2}+W_0^K \sigma_K,\label{W2}
\end{align}
with $\vec\Pi=(\frac{g_e\mu_{B}}{\hbar e}-\frac{1}{2mc})(\frac{e}{c}\vec E + m\dot{\vec v} )$. The first term in Eq.~(2) describes spin-orbit interactions and Thomas precession \cite{Jackson}. The fields $W_0^K$ and   $\vec{W}^{K}$ describe magnetic exchange interactions ($W_0^K$ is the ``Weiss exchange field''), while $\vec \Omega^K$ is the spin connection describing parallel transport of spins in a curved background and interactions with dislocations and disclinations. Note that $\vec v$ and $\frac{1}{\rho}p$ are {\it gauge-invariant} under $U(1)$-gauge transformations. The fields $\vec \Pi$ and $\vec W$ transform {\it homogeneously} under $SU(2)$-gauge transformations, while $\vec \Omega$ transforms {\it inhomogeneously}; ($\vec E$ and $\vec B$ are the electric field and the magnetic induction, respectively).  
All vector quantities in Eqs.~(\ref{W1}) and (\ref{W2}) are expressed in {\it local} coordinate systems. We emphasize that, in the presence of a non-vanishing $SU(2)$-gauge field, the spin current is \textit{not} conserved (it is only covariantly conserved), and, as a consequence, the total magnetization is \textit{not} conserved. 

Introducing the covariant derivatives $D_j = \hbar \partial_j+ia_j+iw_j^K\sigma_K$ and $D_0 = \hbar \partial_t + ia_0 + i w_0^K \sigma_K,$
we can write the Pauli equation in the form
\begin{equation}
iD_0\Psi_t=-\frac{1}{2m\sqrt{g}}\Big(\sum_{i,j} D_i \sqrt{g} g^{ij} D_j\Big)\Psi_t,\label{Pauli}
\end{equation}
where $g^{ij}$ is the metric tensor of the sample background and $g$ its determinant (see Ref.~\cite{FS}).
This equation displays full $U(1)_\text{em}\times SU(2)_\text{spin}$ gauge invariance. It turns out to be the Euler equation corresponding to the action
\begin{align}
S_0(\Psi^{\dagger},\Psi;a,w)=&\int dt \int \sqrt{g} dx \Big[\Psi_{t}^{\dagger}(x)\cdot D_0\Psi_t(x)\nonumber\\
&-\sum_{i,j}\frac{g^{ij}}{2m}\big(D_{i}\Psi_{t}\big)^{\dagger}(x) \cdot D_{j}\Psi_t(x)\Big].
\end{align}
In systems of interacting electrons, a term like
\begin{equation}
S_\text{int}(\Psi^{\dagger},\Psi)\!=\!-\!\int \!\!dt \!\!\int \!\!\sqrt{g}dx \!\!\int \!\!\sqrt{g}dy \vert\Psi_t(x)\vert^{2}\mathcal{U}(x-y)\vert \Psi_t(y)\vert^{2},\nonumber
\end{equation}
where $\mathcal{U}$ is a two-body potential, must be added. Its precise form is not relevant for the following considerations. All that matters is that it is $U(1)_\text{em}\times SU(2)_\text{spin}$ gauge-invariant. The total action $S=S_0+S_\text{int}$ is then manifestly $U(1)_\text{em}\times SU(2)_\text{spin}$ gauge-invariant. 

More generally, one may also gauge ``emergent'' symmetries of electron gases, besides those fundamental symmetries.
In what follows, we mainly focus on electron gases with a bulk mobility gap (insulators), and limit our study to \textit{ground state properties} of such systems, i.e., determine the form of the \textit{effective action} at temperatures $T\approx 0$. For this purpose, we consider the expectation of the propagator, $U_{a,w}(t,s)$, of an interacting electron gas, from time $s$ to time $t$, in the presence of time-dependent gauge fields $a$ and $w$ in the ground state, $\vert\varphi_0\rangle$, of the gas. One defines a ``partition function'' $Z(a,w):=\lim_{\substack{s\rightarrow -\infty\\ t\rightarrow +\infty}}\langle \varphi_0| U_{a,w}(t,s)|\varphi_0\rangle$. This quantity can also be expressed as a functional integral by promoting $\Psi_{t}(x)$ and $\Psi_{t}^{\dagger}(x)$ to Grassmann variables and performing the Berezin integral
$Z(a,w)= \text{const}\cdot\int D\Psi^{\dagger} D\Psi e^{\frac{i}{\hbar}S(\Psi^{\dagger},\Psi; a,w)}$.

The effective action is then defined as
\begin{equation}
S_\text{eff}(a,w)=-i\hbar\ln Z(a,w).\label{Seff}
\end{equation}
It has the following general properties:

{\it(1)} It is the generating function of connected Green functions of the electric- and the spin current densities 
$j^\mu$ and $s^\mu_K$:
\begin{align}
&\frac{\partial S_\text{eff}(a,w)}{\partial a_\mu(x)}=\langle j^\mu(x)\rangle_{a,w}, \hspace{1mm} 
\frac{\partial S_\text{eff}(a,w)}{\partial w_\mu^K(x)}=\langle s^\mu_K(x)\rangle_{a,w},\label{spincurrent}
\end{align}
while higher derivatives yield connected Green functions of current densities. The functional derivative of $S_\text{eff}$ with respect to the metric $g_{ij}$ is the expectation value of the \textit{stress tensor}. (The role of the metric $g_{ij}$ is underemphasized in this Letter; some discussion can be found, e.~g. in Refs.~\cite{FST, Haldane2, Toappear}.) Eqs. (7) are \textit{transport equations}.

{\it (2)} It is gauge invariant:
\begin{align}
&S_\text{eff}(a_\mu+\partial_\mu\chi, Uw_\mu U^{-1}+U\partial_\mu U^{-1})=S_\text{eff}(a_\mu,w_\mu),\nonumber
\end{align}
where $\chi$ is a real-valued function and $U$ denotes a space-time dependent rotation in spin space. We note that electromagnetic gauge invariance and electric current conservation are equivalent, while $SU(2)$-gauge invariance is equivalent to the property that the spin current is covariantly conserved \cite{FS}. 

{\it (3)} Assuming that connected current Green functions have appropriate cluster properties (which is the case for electron liquids with a mobility gap above the ground state energy, such as insulators) and passing to the limit of large distance- and low frequency scales (the \textit{scaling limit}), $S_\text{eff}(a,w)$ can be written as a sum of integrals over {\it local polynomials} in $a$ and $w$ and derivatives thereof. 
These polynomials are gauge-invariant up to total derivatives, which, for samples with non-empty boundaries, may give rise to surface terms depending on the gauge transformation (gauge anomalies). However, since non-relativistic electron gases have an exact $U(1)$- and $SU(2)$- local gauge symmetry, the total effective action of such gases \textit{has} to be invariant under $U(1)$- and $SU(2)$-gauge transformations. It follows that there must exist {\it boundary terms} in the effective action that gauge-transform in such a way as to cancel the gauge-dependent surface terms coming from the bulk effective action. This is called ``anomaly cancellation''. 
In a system confined to a region $\Lambda$ of space-time with non-empty boundary $\partial\Lambda$, the effective action thus has the form
\begin{align}
&S_\text{eff}(a,w)
=\sum_{n} \big(S_{\text{eff},\Lambda}^{(n)}(a,w)+S_{\text{eff},\partial\Lambda}^{(n)}(a |_{\partial\Lambda},w |_{\partial\Lambda}) \big),\nonumber
\end{align}
where $S_{\text{eff},\Lambda}^{(n)}(a,w)$ is a local term of scaling dimension $n$, and 
$S_{\text{eff},\partial\Lambda}^{(n)}(a |_{\partial\Lambda},w |_{\partial\Lambda})$ is a corresponding boundary term canceling an anomaly. 
The boundary terms contain interesting information about degrees of freedom located at the surface of a sample (surface modes). 

{\it (4)} In order to explore physics in the \textit{scaling limit}, it suffices to 
retain the leading (most relevant) terms in the expansion of $S_{\text{eff}}$. 

We thus obtain an {\it explicit} expression for the low-energy effective action, which allows us to make concrete predictions on properties of electron liquids. The fields $a$ and $w$ in $S_{\text{eff}}$ are \textit{classical} and describe internal or external \textit{sources} conjugate to the electric- and spin current density, respectively; the effective action being the generating functional of current Green functions. The action of the low-energy effective \textit{quantum theory} describing the system can usually be inferred from $\text{exp} (iS_{\text{eff}})$ by functional Fourier transformation; see \cite{FST, FKST, WZ, Toappear}.

While the starting point of \textit{Landau theory} is the identification of global symmetries of a system and of a space of order parameters on which these symmetries act, along with the allowed patterns of spontaneous symmetry breaking \cite{Leutw1, Leutw2}, the starting point of our ``gauge theory of states of matter" is the idea that global transformations acting on the degrees of freedom of an electron gas, which, in the absence of gauge fields, describe (fundamental or emergent) symmetries of the system, shall be gauged, and the response of the system to turning on corresponding gauge fields be analyzed (Eq.~(\ref{spincurrent})). This enables one to identify states of matter exhibiting ``topological order". In the following, we sketch a variety of examples (some quite familiar, others new) of applications of the strategy described in {\it (1)} through {\it (4)} to electron gases.

{\it $(2+1)D$ examples}. (i) We start by considering a two-dimensional electron gas subject to a strong, uniform magnetic field $\vec{B}^{(0)}=\nabla \wedge \vec{A}^{(0)}$ perpendicular to the sample surface (and defining the local $z$-axis). The fields $w$ and $\vec v$ are set to zero. We assume that, for appropriate choices of the external field $\vec{B}^{(0)}$, the bulk Hamiltonian of the gas has a mobility gap above the ground state energy, with the spins of the electrons parallel or anti-parallel to $\vec{B}^{(0)}$. We propose to study the response of the electron gas to small fluctuations in the electromagnetic field and in the curvature of the sample surface. The total electromagnetic vector potential is denoted by $\vec{A}^{\text{em}}=\vec{A}^{(0)}+\vec{A}$. If the sample surface of the gas has non-vanishing Gauss curvature, the $U(1)$-gauge field $a_{\mu}$  contains a contribution describing parallel transport on the sample surface (rotations around the local $z$-axes), besides $\vec{A}$. Equation~(\ref{Seff}) leads to  
\begin{align}
&S_\text{eff}(a)=\frac{\sigma_H}{2}\int_{\Lambda} dt dx \varepsilon^{\mu\nu\rho} a_\mu\partial_\nu a_\rho+\Gamma(a|_{\partial\Lambda})\nonumber\\
&=\frac{\sigma_H}{2}\int_{\Lambda} dt dx \left[\varepsilon^{\mu\nu\rho}A_{\mu}\partial_{\nu}A_{\rho} + \frac{2}{e}A_{0}K\right] + \Gamma(a|_{\partial\Lambda}),\label{S_QHE}
\end{align}
where $\sigma_H$ is the Hall conductivity, $\varepsilon^{\mu\nu\rho}$ the Levi-Civita anti-symmetric tensor, ${A_0=\varphi}$ the scalar potential, and $K$ is the Gauss curvature of the sample. This action describes the \textit{Quantum Hall Effect} (QHE): 
$j^{\mu} = \sigma_H \left[\varepsilon^{\mu\nu\rho} \partial_{\nu} A_{\rho} + e^{-1}K \delta^{\mu 0}\right]$. 
Curvature effects in the context of the QHE have first been described in Refs.~\cite{FS, FST} (see also Ref.~\cite{WZ}), and have recently attracted renewed interest (see, e.g., Ref.~\cite{Haldane2}). The edge action $\Gamma(a|_{\partial\Lambda})$ is the well known anomalous chiral action in $(1+1)$ dimensions, which cancels the gauge anomaly of the first (bulk) term  (the $(2+1)$-dimensional Chern-Simons action) in Eq.~(\ref{S_QHE}). It is the generating function of Green functions of  chiral electric edge currents propagating along the boundary $\partial\Lambda$ of the sample \cite{footnote3c}, first discovered in Ref.~\cite{Halperin}; (see also \cite{Wen, FK}). 
A similar analysis applies to layers of atomic quantum gases if one replaces $\vec{A}^\text{em}$ by the velocity field, $\vec{v}$, corresponding to a rapidly rotating sample. This led to the prediction \cite{FST} of a Hall effect in rotating Bose gas layers confined by harmonic traps canceling the centrifugal forces; see also \cite{CWG}.

If parity and time reversal are symmetries of the system then the Chern-Simons terms in Eq.~(\ref{S_QHE}) cannot appear. For a two- or three-dimensional insulator, the leading term is then given by $ \int dt dx \epsilon \vec E^2 - \int dt dx (1/\mu) \vec B^2\nonumber$, where $\epsilon$ is the dielectric constant and $\mu$ the magnetic permeability.
If one drops the assumption that the system has a bulk mobility gap, i.e., is an insulator, then a term
$(2\lambda^{2})^{-1}\int dt dx(\vec{a}^T)^2$ may appear in the effective action, which describes an ordinary superconductor. Here $\vec{a}^T$ is the transverse part of $\vec{a}$ and $\lambda$ the London constant of the superconductor. For two-dimensional systems, there is an intriguing duality involving the low-energy effective theory, which maps insulators to superconductors, and conversely, and incompressible Hall fluids to themselves; see Ref.~\cite{FST}.

(ii) Next, we introduce an $SU(2)$-gauge field, $w$, describing spin-orbit- and/or exchange-interactions, but, for simplicity, 
omit the $U(1)$-gauge field $a$. 
The effective action of two-dimensional gapped magnetic systems in the scaling limit is then given by \cite{FS, FST}
\begin{align}
&S_\text{eff}(w)=\chi\int dt dx \text{Tr} (w_0)^2\nonumber +\tilde{\chi}\int dt dx \text{Tr} (\vec{w} - \vec{\Omega})^2\nonumber\\
&+\frac{k}{4\pi}\int dt dx \varepsilon^{\mu\nu\rho} \text{Tr} \left(w_\mu\partial_\nu w_\rho+\frac{2}{3}w_\mu w_\nu w_\rho \right)\nonumber\\
& +\text{edge action depending on $w|_{\partial\Lambda}$},
\label{S_2d}
\end{align}
up to ``irrelevant" terms.

The coefficient $\chi$ is proportional to the magnetic susceptibility of a paramagnetic insulator. The second term on the right side 
describes the \textit{spin Hall effect} \cite{FS, FST, Kane2005, Bernevig2006, Koenig2007, Koenig2008}: We consider a 
two-dimensional electron gas in the $x$-$y$ plane of physical space 
in the presence of 
a strong intrinsic electric field, $E_{\bot}$, supported near the boundary of the sample and perpendicular to it; ($E_{\bot}$ confines the electrons to the sample). According to Eq.~(\ref{W1}), this electric field corresponds to an $SU(2)$-gauge field, $\vec{w}$, \textit{parallel} to the sample's boundary and supported near it, which describes spin-orbit interactions. It is given by 
$w_\parallel = \text{const}\cdot E_{\bot} \sigma _{z}$; all other components of $\vec{w}$ vanish, and $\vec{\Omega} = 0$.
According to Eq.~(\ref{spincurrent}), the response equation resulting from the second term on the right side of Eq.~(\ref{S_2d}) is 
\begin{align}
\langle s^{3}_{\parallel}\rangle = \tilde{\chi} E_{\bot},
\end{align}
while all other components of the spin current vanish. This equation describes the spin Hall effect; 
(see also~Ref.~\cite{Grafetal} for mathematically rigorous results).

The third term on the right hand side of Eq.~(\ref{S_2d})  is an $SU$(2)-Chern-Simons term that has a gauge anomaly at the boundary of the sample and requires the addition of an ``edge action" canceling this anomaly. 
The coefficient $k$ must be an \textit{integer} \cite{DesJackTempl,Witten}. The edge action canceling the anomaly of this term is the generating function of Green functions of chiral current operators generating an $SU(2)$ current algebra at level $k$. The irreducible unitary representations of an $SU(2)$ current algebra at level $k$ are labeled by a spin quantum number $s=0, \frac{1}{2}, \ldots \frac{k}{2}$. Level $k=0$ corresponds to insulators \textit{without} chiral boundary currents, while $k\geq1$ describes an interesting type of topological insulator with \textit{chiral edge spin currents} carried by quasi-particles of, typically, spin $\frac{1}{2}$. 
Chiral edge spin currents can be expected to be present in gapped systems with broken time reversal, such as certain quantum Hall fluids and graphene-like systems (with pairs of two-component Dirac fermions with spin $\frac{1}{2}$ and a small mass as bulk quasi-particles); see Refs.~\cite{Mudry, Haldane} and references given therein.

{\it $(3+1)D$ examples} (i) We first limit our considerations to insulators, i.e., materials with a bulk gap, and determine their effective actions in the presence of $U(1)$- and $SU(2)$-gauge fields. For simplicity, we set $\vec v=0$. The effective action then reads
\begin{align}
&S_\text{eff}(a,w)=\frac{1}{2}\int_\Lambda dtdx \left[\epsilon \vec E^2-(1/\mu)\vec B^2+\frac{\gamma}{2\pi^{2}} \vec E\cdot\vec B\right]\nonumber\\
&+\frac{1}{2}\int_\Lambda dtdx \text{Tr}\Bigg[ \epsilon_{w}\sum_{i=1}^{3}(F_w)_{0i}^{2} - ({1/\mu_{w}})\sum_{i,j=1}^3(F_w)_{ij}^{2}\nonumber\\
    &+\frac{\theta}{16\pi^2} \varepsilon^{\mu\nu\rho\sigma} (F_w)_{\mu\nu}(F_w)_{\rho\sigma}\Bigg]+\text{less relevant terms}.\label{Seff_topins}
\end{align}
The third term in Eq.~(\ref{Seff_topins}) is a topological term; (it is really a surface term). For general values of    $\gamma$, it breaks parity and time reversal symmetry, except when  $\gamma = 0,\pi$.
The $SU(2)$-field strength $F_{w}$ is given by $(F_w)_{\mu\nu} = \partial_{\mu}w_{\nu}-\partial_{\nu}w_{\mu} + \frac{i}{2}[w_{\mu},w_{\nu}].$ We note that, because $\frac{1}{32\pi^2}\int dt dx\text{Tr}[\varepsilon^{\mu\nu\rho\sigma}(F_w)_{\mu\nu}(F_w)_{\rho\sigma}]$
is an integer, parity invariance of the bulk implies that $\theta   =0 $ or $ \pi$, and conversely, similarly as for $\gamma$.
We assume that the sample $\Lambda$ has the geometry of a ``slab'' and denote its boundary by $\partial\Lambda$. From Stokes' theorem we find that 
\begin{align}
& \int_\Lambda dtdx \vec E\cdot\vec B = \frac{1}{2} \int_{\partial\Lambda} dtdx \varepsilon^{\mu\nu\rho}a_\mu\partial_\nu a_\rho=: 2\pi^{2} \Gamma_{\partial \Lambda}(a),\\
&\frac{1}{4}\int_\Lambda dtdx \varepsilon^{\mu\nu\rho\sigma}  \text{Tr}[(F_w)_{\mu\nu}(F_w)_{\rho\sigma}]\nonumber\\
&=\int_{\partial \Lambda}dtdx \varepsilon^{\mu\nu\rho} \text{Tr}\Big[w_\mu \partial_{\nu}w_{\rho}+ \frac{2}{3} w_\mu w_\nu w_\rho\Big]=: 4\pi^{2} \Gamma_{\partial \Lambda}(w).\label{CS}
\end{align}
The boundary term given by $\pi\Gamma_{\partial\Lambda}(a)$ is the effective action of a \textit{charged, 2-component Dirac fermion} \cite{DesJackTempl, footnote5}. 
Thus, $\gamma/\pi$ determines the number of species of charged Dirac fermions propagating along the boundary. 
On the right hand side of Eq.~(\ref{CS}) we recognize the {\it non-abelian Chern-Simons term}: $ \pi\Gamma_{\partial\Lambda}(w)$ is the effective action of a 2D {\it relativistic fermion with ``$SU(2)$ isospin"}. We conclude that the gauge-invariant action (\ref{Seff_topins}) predicts {\it topological insulators}, for $\gamma$ and/or $\theta$ $\neq 0$; ($\gamma, \theta=0$ corresponds to ordinary insulators). 

If we set $w=0$ and add a term $\frac{1}{2\lambda^{2}}\int dtdx(\vec{a}^T)^2$ in Eq.~(\ref{Seff_topins}) we obtain the effective action describing topological superconductors for $\gamma=\pi$ (while $\gamma=0$ corresponds to ordinary superconductors). 
In fact, $S_\text{eff}=\frac{1}{2\lambda^{2}}\int dtdx(\vec{a}^T)^2+\frac{\gamma}{32\pi^2}\int dtdx \varepsilon^{\mu\nu\rho\sigma} F_{\mu\nu} F_{\rho\sigma}$ and Eq.~(\ref{spincurrent}) lead to the {\it London equation} 
\begin{equation}
-\frac{e^2}{mc}n (\vec{A}^\text{em})^T=\vec{j},
\end{equation}
where $\lambda^{2}  = - mc/e^{2}n$, with $n$ the condensate density.

In $p$-wave superconductors, electron pairs have spin $1$, and the
$SU(2)$-gauge field $w$ can no longer be neglected in the effective
action. The introduction of $w$-dependent terms in the effective
action yields a description of novel topological $p$-wave
superconductors.

It is of considerable interest to observe that a system may
respond to switching-on an external magnetic field that increases
its free energy by setting itself into motion, thus generating
a current or velocity field that (partially) cancels the
external gauge fields; (generalized ``Lenz principle").
Examples of this phenomenon include the Meissner-Ochsenfeld effect (super-current canceling an external electromagnetic vector potential) and the Einstein-deHaas-Barnett effect (velocity field whose vorticity cancels an external magnetic field). Vortices in superfluid helium can also be studied from this point of view, in close analogy to Abrikosov vortices in a superconductor, but with $\vec{A}$ replaced by $\vec{v}$; see \cite{Volovik, FST}. Similarly, the effect of sonoluminescence can be understood as an electromagnetic  response offsetting pressure- and velocity fluctuations in a gas. 

(ii) {\it Axion electrodynamics}. If we promote the coupling $\gamma$ in Eq.~(\ref{Seff_topins}) to a dynamical variable (field) $\phi$, i.e., replace $\frac{\gamma}{4\pi^2}\int dtdx  \vec E\cdot\vec B=\frac{\gamma}{32\pi^2}\int dtdx \varepsilon^{\mu\nu\rho\sigma} F_{\mu\nu}F_{\rho\sigma}$ by
\begin{align}\label{ax}
&\frac{1}{32\pi^2}\int dtdx \varepsilon^{\mu\nu\rho\sigma} (\gamma+l\phi)F_{\mu\nu}F_{\rho\sigma},
\end{align}
where $l$ is a parameter with the dimension of a length, 
and add the term
$\frac{1}{\alpha} \int dtdx [\frac{1}{2}\partial_\mu\phi\partial^\mu\phi+U(\phi)]$,
then we obtain an action containing a coupling of electrons to an axion field; see Refs.~\cite{FP,Werner2000}. The axion potential $U(\phi)$ is usually periodic in $\phi$ with period $2\pi/l$. If $U\neq0$ then our theory predicts the existence of domain walls across which the value of the axion field changes by an integer multiple of $2\pi/l$ . These domain walls, which, for entropic reasons, must occur in the bulk of axionic topological insulators (or axionic topological superconductors),  support massless (i.e., extended) charged modes. 

{\it $(4+1)D$ examples}. The QHE has a $(4+1)$-dimensional cousin first studied in Refs.~\cite{FP, Werner2000}. Let us consider a five-dimensional system confined to a slab $0<x^4<L$ consisting of very heavy charged four-component Dirac fermions. If these Dirac fermions are coupled to an external electromagnetic vector potential $A^{(5)}$ and are then integrated out, the effective action, as given by Eq.~(\ref{Seff}), becomes
\begin{equation}
S_\text{eff}(A^{(5)})=S_{EM}^{(5)}(A^{(5)})-S_{CS}^{(5)}(A^{(5)})+\Gamma_{\partial\Gamma}(A^{(5)}|_{\partial\Gamma}),\nonumber
\end{equation}
with $S_{EM}^{(5)}(A^{(5)})=-\frac{1}{4L\alpha}\int dtdx (F^{(5)})^{\mu\nu}(F^{(5)})_{\mu\nu}$ 
the $(4+1)$-dimensional analogue of the Maxwell action ($F^{(5)}_{\mu\nu}=\partial_\mu A^{(5)}_\nu - \partial_{\nu} A^{(5)}_{\mu}$), 
and $S_{CS}^{(5)}(A^{(5)})$ proportional to the five-dimensional Chern-Simons action
\begin{equation}
S_{CS}^{(5)}(A^{(5)})=\frac{N}{96\pi^2}\int_\Lambda dtdx \epsilon^{\mu\nu\delta\rho\epsilon}A^{(5)}_\mu F^{(5)}_{\nu\delta} F^{(5)}_{\rho\epsilon},
\label{S_CS}
\end{equation}
where $N=1,2,\ldots$ is the number of fermion species.
The boundary term $\Gamma_{\partial\Gamma}((A^{(5)}|_{\partial\Gamma})$ must be introduced in order to ensure the gauge invariance of the effective action in the slab geometry. It describes {\it massless chiral fermions} on the $(3+1)$-dimensional ``top'' and ``bottom'' boundary components (or ``branes") of the slab. These chiral fermions may acquire a mass through tunneling between the two boundary components. 
Equation~(\ref{spincurrent}), 
applied to the Chern-Simons action (\ref{S_CS}), yields the $(4+1)D$ analogue of Hall's law, 
$j^\mu=\frac{N}{32\pi^2}\epsilon^{\mu\nu\delta\rho\epsilon}F^{(5)}_{\nu\delta} F^{(5)}_{\rho\epsilon}$.
This equation, together with the conservation of the total current, $j^\mu_\text{tot}=j^\mu_\text{bulk}+j^\mu_\text{brane}$, reproduces the so-called chiral anomaly in $(3+1)D$: $\partial_\mu j^\mu_\text{brane}=\sigma_H \vec E\cdot \vec B$, $\sigma_{H} = N/{4\pi^{2}}$.
The chiral anomaly implies the relation $\vec{j}_\text{em} = -\sigma_H (\mu_{L} - \mu_{R}) \vec{B}$, where $\mu_{L,R}$ are chemical potentials corresponding to left- and right-handed fermions \cite{Alekseev1998}.

Axion electrodynamics in $(3+1)D$ can be recovered from the $(4+1)D$ theory discussed here by dimensional reduction \cite{FP, Werner2000}: Suppose that the five-dimensional electromagnetic field is $x^4$-independent. Then $\frac{1}{L}\int_{\gamma^{(4)}} dx^{4} A_4$, with $\gamma^{(4)}$ a curve parallel to the $x^{4}$-axis from one boundary component to the other one, plays the role of the axion field, $\phi$, in the $(3+1)D$ action of axion QED, and the thickness, $L$, of the ``slab" in $(4+1)$-dimensional space-time is related to the parameter $l$ in front of the axion term in Eq. (\ref{ax}). The $(4+1)D$ formulation shows that the time-derivative of the axion field plays the role of a (now space-time dependent) {\it chemical potential difference} between left- and right-handed fermions: $-L\partial_0\phi=-LE_4=\mu_L-\mu_R$ \cite{FP,Werner2000,BoyFrRuch}.

{\it Instabilities.} The equations of motion derived from the Maxwell action combined with Eq.~(\ref{ax}), namely the Maxwell equations for the electromagnetic field and the equation 
\begin{equation}
\partial^\mu\partial_\mu\phi=-\frac{l\alpha}{4\pi^2}\vec E\cdot \vec B-U'(\phi),\label{eom_axion}
\end{equation}
exhibit various instabilities that can be discovered by linearizing the equations of motion around special solutions. The first such instability was identified in Refs.~\cite {FP, Werner2000}, (see also \cite{BoyFrRuch, BRShap}). If the equations are linearized around $\vec{E}=\vec{B}=0$ and a non-trivial, \textit{spatially constant} solution of the equation $\partial^\mu\partial_\mu \phi=-U'(\phi)$, (e.~g., $\dot{\phi} = \text{const} \neq 0$, for $U(\phi) \equiv 0$), one finds unstable Fourier modes of the electromagnetic field, for small wave vectors, describing the generation of rather homogeneous helical magnetic fields that may be relevant in cosmology \cite{footnote4b}. A related instability of more direct interest in condensed matter physics has been analyzed in Ref.~\cite{Ooguri}. Here, the equations of motion for a compact sample of an axionic topological insulator are linearized around a constant electric field $\vec{E}\neq0$, with $\vec{B}=0, \phi=0$. If the electric field applied to the sample exceeds some critical strength then, in the interior of the sample, it is screened by surface charges, and a non-zero magnetic field is generated inside the sample.

\textit{Conclusions.} The key idea explored in this Letter is to promote global transformations acting on the degrees of freedom of systems of condensed matter, which, in the absence of gauge fields, correspond to fundamental or emergent global symmetries, to local gauge transformations, with the purpose of studying the response of such systems to turning on the corresponding gauge fields. These gauge fields \textit{may}, but \textit{need not} correspond to \textit{physical} fields. They are introduced in order to encode transport equations of the system. By using very general principles, in particular gauge invariance, anomaly cancellation, cluster properties and power counting, one is able to determine the general form of the effective action of such systems in the scaling limit. 
This leads to a partial classification of states of condensed matter, including ``topological phases", and of the corresponding surface states. 

\acknowledgments
We thank the referee for very helpful comments. JF thanks R. Morf, his former collaborators T. Kerler, B. Pedrini, U. M. Studer, E. Thiran, and his recent collaborators A. Boyarsky, I. Levkivskyi, O. Ruchayskiy, E. Sukhorukov, for numerous useful discussions. He thanks D. Haldane, S. Sachdev and E. Witten for interesting comments. The work of JF is supported by `The Fund for Math' and `The Monell Foundation' of the IAS.


\begin{thebibliography}{99}
\bibitem{footnote1} That the velocity field appears as a contribution to the $U(1)$-vector potential can be understood by recalling that the Lorentz- and the Coriolis force have the same form.
\bibitem{footnote3c} Effects of the motion of the sample harboring the electron gas can be accounted for by adding the velocity field $\vec{v}$ describing the motion of the sample to the $U(1)$-connection \cite{FS,FST}.
\end{thebibliography}
\end{document}